\documentclass[11pt,titlepage]{article}
\usepackage{latexsym,graphicx,amsthm,amssymb,bm,url}
\usepackage{secdot}
\usepackage{color}
\usepackage{bbm}

\usepackage[a-1b]{pdfx} 
\usepackage{hyperref}

\usepackage{natbib}
 \bibpunct[, ]{(}{)}{,}{a}{}{,}%

\theoremstyle{definition}

\newtheorem{example}{Example}

\newcommand{\E}{\ensuremath{\mathbb{E}}}

\begin{document}


\begin{center}
{\LARGE Quadratic Hedging and Optimization of Option Exercise Policies}
\end{center}



\begin{center}
  {\large
    \mbox{Nicola Secomandi}
  }\\
  {
   \mbox{Tepper School of Business, Carnegie Mellon University, Pittsburgh, PA, USA, ns7@andrew.cmu.edu}
  }
\end{center}


\begin{center}
December 2019; Revised: October 2021
\end{center}




\begin{center}
{\small {\bf Abstract}}
\end{center}

{\small{\noindent
Quadratic hedging of option payoffs generates the variance optimal martingale measure.
When an option features an exercise policy and its cash flows are hedged according to this approach, it may be tempting to optimize such a policy under this measure.
Because the variance optimal martingale measure may not be an equivalent probability measure, focusing on American options we show that the resulting exercise policy may be unappealing.
This drawback can sometimes be remedied by imposing time consistency on exercise policies, but in general persists even in this case, which compounds the familiar issue that valuing an option using this measure may not result in an arbitrage free value.
An alternative and known approach bypasses both of these pitfalls by optimizing option exercise policies under any given equivalent martingale measure and anchoring quadratic hedging to the resulting value of this policy.
Additional research may assess on realistic applications the magnitude of the limitations associated with optimizing option exercise policies based on the variance optimal martingale measure.
}

\baselineskip 20pt plus .3pt minus .1pt

\section{Introduction}

American and related options play important roles in financial engineering (\citealp{kn:Shiryaev1999,kn:Duffie2001,kn:BinghamKiesel2004,kn:Shreve2004}, and~\citealp{kn:Detemple2005}) and real options (\citealp{kn:DixitPindyck1994,kn:Trigeorgis1996,kn:Guthrie2009}, and~\citealp{kn:SecomandiSeppi2014}).
They are traded in several stock, commodity, and energy exchanges and form the building blocks of models that represent the managerial flexibility embedded in projects, e.g., buying or selling a traded asset at a given price, building a new plant, developing land or a new technology, pausing and resuming production, and managing energy and commodity infrastructure and conversion facilities on a merchant basis.
A key feature of these options is their exercise policy.

Market incompleteness is the norm in both theory and practice (see, e.g.,~\citealp{kn:Shiryaev1999,kn:BinghamKiesel2004,kn:ContTankov2004,kn:Shreve2004,kn:BenthEtAl2008},~\citealp[Part III]{kn:Swindle2014}, \citealp{kn:Mahoney2016}, and~\citealp{kn:Swindle2016}).
There are various approaches to hedge option payoffs in incomplete markets (see, e.g.,~\citealp[Ch. 7]{kn:BinghamKiesel2004},~\citealp[Ch. 10]{kn:ContTankov2004}, \citealp{kn:Staum2008}, and \citealp{kn:RheinlanderSexton2011}).
Quadratic hedging is a practical method based on approximate replication of option cash flows based on self financing trading policies (see, e.g., \citealp{kn:Schal1994,kn:Schweizer1995,kn:Schweizer1996,kn:BertsimasEtAl2001}, and~\citealp{kn:Cerny2004}).
Even if typically used for European options, for which the option exercise policy is given, a simple adaptation of the model of~\citet{kn:SecomandiYang2021} makes it possible to apply quadratic hedging to options for which an exercise policy needs to be determined.
In particular, \citet{kn:SecomandiYang2021} extend the conditional quadratic hedging proposal of~\citet{kn:Secomandi2021} for assets with a single cash flow on a fixed date, such as European options, to assets distinguished by streams of cash flows, of which American options are a special case.

Quadratic hedging generates the so called variance optimal martingale measure and sets the optimal initial capital position of the hedge equal to the expectation of the discounted option cash flows taken under this measure.
If we take this expectation as a proxy, in general, for the value of an option, one may be tempted to optimize an option exercise policy by maximizing this quantity.
Focusing, for expositional simplicity, on American options on a single commodity futures in discrete time, based on the work of~\citet{kn:Secomandi2019} for assets with cash flows of European type, we show that doing so can yield unappealing exercise policies if the variance optimal martingale measure is not an equivalent probability measure, that is, a risk neutral one.
Although restricting attention to time consistent policies can sometimes alleviate this deficiency, it cannot always eliminate it.
Further, a known potential drawback~\citep{kn:Schweizer1995,kn:Schweizer1996} of the variance optimal martingale measure is that it can fail to assign an arbitrage free value to an option.
These issues can be avoided by (i) optimizing an option exercise policy under any given risk neutral measure, as usual (see, e.g.,~\citealp[Chapter 8]{kn:Glasserman2004} and~\citealp{kn:Detemple2005}), and (ii) anchoring quadratic hedging to the resulting value of this policy for its choice of the initial capital position, which drives the trading of the financial instruments, as done by~\citet{kn:SecomandiYang2021}.
In spite of the shortcomings of exercise policy optimization under the variance optimal martingale measure, given that standard quadratic hedging yields minimal mean square replication error further research may measure their scale in practical applications.

Section~\ref{sc:model} introduces the model.
We discuss quadratic hedging of a given exercise policy in Section~\ref{sc:qh} and deal with the optimization of such a policy in Section~\ref{sc:opt}.
Section~\ref{sc:alternative} presents the stated alternative approach.
We conclude in Section~\ref{sc:concl}.

\section{Model}
\label{sc:model}

Consider an American option on a commodity futures that can be exercised at one of $I$ dates with respective indices in set $\mathcal{I}:=\{0,1,\ldots,I-1\}$.
The futures price on date $T_i$ is $F_i\in\mathbb{R}_+$.
Exercising the option on date $T_i$ when the futures price is $F_i$ yields the cash flow $C_{i}(F_{i})$.
For convenience, we assume that this cash flow is nonnegative, e.g., it is $\max\{F_i-K,0\}$ for a call option with strike price $K$, and the option is always exercised on the last date, $T_{I-1}$, if it is still alive at this time, in which case the cash flow $C_{I-1}(F_{I-1})$ may be zero for some values of the futures price $F_{I-1}$.

A known Markovian stochastic process governs the evolution of the futures price.
It is independent of the option exercise decisions.
There is a risk free bond.
Its associated discount factor from date $T_{i}$ back to date $T_{i-1}$ with $i\in\mathcal{I}\setminus\{0\}$ is $D_{i}$.
The dynamics of the price of the futures give rise to an incomplete market.
That is, in general the cash flows of feasible option exercise policies cannot be perfectly replicated by dynamically trading this contract and a risk less bond.
There are no arbitrage opportunities.

Let $\pi$ be a feasible option exercise policy.
Its decision rule $X^\pi_i(F_i)$ prescribes the date index and futures price pair $(i,F_i)$ for which to exercise the option.
It equals one if the option is exercised and zero otherwise.
The set of feasible policies is $\Pi$.
Denote by $\iota^\pi$ the stopping index corresponding to the date policy $\pi$ exercises the option, that is, $T_{\iota^\pi}$ is the corresponding stopping time.
By assumption, if the option has not been exercised by date $T_{I-1}$ then $\iota^\pi$ equals $I-1$. 

\section{Quadratic Hedging for a Given Exercise Policy}
\label{sc:qh}

This section adapts to the setting of this paper the conditional quadratic hedging model for streams of cash flows of~\citet{kn:SecomandiYang2021} by optimizing, as in standard quadratic hedging~(\citealp{kn:Schweizer1995,kn:Schweizer1996,kn:BertsimasEtAl2001,kn:Cerny2004}, and~\citealp{kn:Secomandi2019}), the initial capital position (the risk less bond position in this paper).


Fix an option exercise policy $\pi\in\Pi$.
Let $\psi$ be a self financing financial trading policy and $\Psi$ be the set of all such policies.
If the policy $\pi$ does not immediately exercise the option, that is, $\iota^\pi \neq 0$, each policy $\psi$ takes positions in the risk less bond and the futures on date $T_0$ and adjusts them on each subsequent date $T_1$ through $T_{\iota^\pi-1}$, that is, trading stops when the option is exercised.
Otherwise, the bond position is taken on the initial date and no futures is ever traded.


Denote by $V^\psi_{i}$ the value of the portfolio of policy $\psi$ on date $T_i$ with $i=0,\ldots,\iota^\pi$.
If $\iota^\pi \neq 0$, let $\Theta_{i}(\pi,V^\psi_i,F_i)$ be the futures trading decision rule of policy $\psi$ on date $T_i$ with $i<\iota^\pi$.
On date $T_0$ the portfolio value is the value of the bond position chosen by policy $\psi$, $V^\psi_0$, and if $\iota^\pi\neq 0$ on each later date $T_{i}$ with $i\leq\iota^\pi$ it equals
\begin{equation}
V^\psi_{i} = \frac{V^\psi_{i-1}}{D_{i}} + \Delta F_{i} \Theta_{i-1}(\pi,V^\psi_{i-1},F_{i-1}),
\label{eq:msf}
\end{equation}
where we define $\Delta F_{i}$ as $F_{i}-F_{i-1}$ and the second term of the sum in~(\ref{eq:msf}) is the mark to market cash flow on the futures position created on date $T_{i-1}$.

Let $D_{i,I-1} :=\prod_{j=i+1}^{I-1} D_j$ for each $i\in\mathcal{I}\setminus\{I-1\}$ and $D_{I-1,I-1}:=1$.
The quantity
\begin{equation}
D^{-1}_{\iota^\pi,I-1}\cdot\left(C_{\iota^\pi}(F_{\iota^\pi})-V^{\psi}_{\iota^\pi}\right)
\label{eq:error}
\end{equation}
is the error expressed in date $T_{I-1}$ dollars incurred by policy $\psi$ when replicating the cash flow of the exercise policy $\pi$.
Denote by $\mathbb{E}$ expectation under the statistical measure.
Quadratic hedging aims at finding a policy $\psi$ with minimal mean squared replication error:
\begin{equation}
\min_{\psi\in\Psi}\E\left[D^{-2}_{\iota^\pi,I-1}\cdot\left(C_{\iota^\pi}(F_{\iota^\pi})-V^{\psi}_{\iota^\pi}\right)^2\mid F_0\right].
\label{eq:hedgemod}
\end{equation}
The date $T_{I-1}$ value of the profit and loss (P\&L) from purchasing the option at $V^\psi_0$ on date $T_0$ and exercising it by following policy $\pi$, that is, the unhedged P\&L, is $D^{-1}_{\iota^\pi,I-1}C_{\iota^\pi}(F_{\iota^\pi})-D^{-1}_{0,I-1}V^\psi_0$.
Expression~(\ref{eq:error}) equals
\[
D^{-1}_{\iota^\pi,I-1}C_{\iota^\pi}(F_{\iota^\pi})-D^{-1}_{0,I-1}V^{\psi}_0-\sum_{i=1}^{\iota^\pi}D^{-1}_{i,I-1}\Delta F_{i} \Theta_{i-1}(\pi,V^\psi_{i-1},F_{i-1}),
\]
which can be interpreted as the date $T_{I-1}$ value of supplementing the unhedged P\&L with the cash flows corresponding to trading the futures according to policy $-\psi$, which yields the hedged P\&L associated with this policy.
Hence, model~(\ref{eq:hedgemod}) determines the initial risk free bond position and the futures trading policy that minimize the second moment of the hedged P\&L.

Model~(\ref{eq:hedgemod}) can be solved based on stochastic dynamic programming, similar to the development in~\citet{kn:BertsimasEtAl2001,kn:Cerny2004,kn:Secomandi2019}, and~\citet{kn:SecomandiYang2021}.
Financial trading occurs only if the option is alive.
Thus, we formulate the stochastic dynamic program under this assumption.
The value function in stage $i\in\mathcal{I}$ and state $(V_{i},F_i)\in\mathbb{R}\times \mathbb{R}_+$ satisfies
\[
J^\pi_i(V_i,F_i) =
\left\{
\begin{array}{ll}
  D_{i,I-1}^{-2}\cdot\left(C_i(F_i)-V_{i}\right)^2, & \mbox{if } X^\pi_i(F_i) =  1 \mbox{ or } i=I-1,\\
\min_{\theta_i\in\mathbb{R}} \E\left[J^\pi_{i+1}\left(
V_i/D_{i+1}+\Delta F_{i+1}\theta_i, F_{i+1}\right)
 \mid V_i,F_i\right], & \mbox{if } X^\pi_i(F_i) = 0 \mbox{ and } i\neq I-1.
\end{array}
\right .
\]

For each stage $i\in\mathcal{I}$ and futures price $F_{i}\in\mathbb{R}_+$ such that $X^\pi_i\left(F_i\right) = 1$ or $i=I-1$ let $a^\pi_{i}(F_{i}) := D_{i,I-1}^{-2}$, $b^\pi_{i}(F_{i}) := C_i(F_{i})$, and $c^\pi_{i}(F_{i}) := 0$.
For each stage $i\in\mathcal{I}\setminus\{I-1\}$ and futures price $F_i\in\mathbb{R}_+$ such that $X^\pi_i\left(F_i\right) = 0$ define
\begin{eqnarray*}
q^\pi_{i}(F_{i}) &:=& 
\frac{
\E\left[a^\pi_{i+1}(F_{i+1})\Delta F_{i+1}\mid F_{i}\right]
}
{
\E\left[a^\pi_{i+1}(F_{i+1}) (\Delta F_{i+1})^2 \mid F_{i}\right]
},\\
a^\pi_{i}(F_{i}) &:=& \frac{1}{D^2_{i+1}}\E\Big[a^\pi_{i+1}(F_{i+1}) (1- q_{i}(F_{i}) \Delta F_{i+1})^2\mid F_{i}\Big],\\
p^\pi_{i}(F_{i}) &:=&
\frac
{
\E\left[a^\pi_{i+1}(F_{i+1})b^\pi_{i+1}(F_{i+1})\Delta F_{i+1}\mid F_{i}\right]
}
{
\E\left[a^\pi_{i+1}(F_{i+1})(\Delta F_{i+1})^2\mid F_{i}\right]
 },\\
b^\pi_{i}(F_{i}) &:=& \frac{1}{a^\pi_{i}(F_{i})D_{i+1}}
\E\left[a^\pi_{i+1}(F_{i+1})
(b^\pi_{i+1}(F_{i+1})-p^\pi_{i}(F_{i})\Delta F_{i+1})
(1- q_{i}(F_{i})\Delta F_{i+1})\mid F_{i}\right],\\
c^\pi_{i}(F_{i}) &:=& \E\left[c^\pi_{i+1}(F_{i+1})\mid F_{i}\right]
+\E\left[a^\pi_{i+1}(F_{i+1})
(b^\pi_{i+1}(F_{i+1})
- p^\pi_{i}(F_{i})\Delta F_{i+1})^2\mid F_{i}\right]\nonumber\\
&&-a^\pi_{i}(F_{i}) \left[b^\pi_{i}(F_{i})\right]^2.
\end{eqnarray*}
The value function $J^\pi_{i}(V_{i},F_{i})$ can be written as
\[
a^\pi_{i}(F_{i}) \left[b^\pi_{i}(F_{i})-V_{i}\right]^2 +c^\pi_{i}(F_{i}).
\]
The term $a^\pi_{i}(F_{i})$ is nonnegative.
The optimal futures trading decision rule is
\[
p^\pi_{i}(F_{i}) - \frac{1}{D_{i+1}} q^\pi_{i}(F_{i}) V_{i}.
\]
The optimal initial bond position, which minimizes the function $J_0(\cdot,F_0)$, is $b^\pi_{0}(F_{0})$.

It is useful to characterize each term $b^\pi_i(F_i)$.
Let $\mathbb{E}^{\mathrm{VO}(\pi)}\left[\cdot\mid \iota^\pi>i,F_i\right]$ be expectation under the variance optimal martingale measure (see, e.g., \citealp{kn:Schweizer1995,kn:Schweizer1996}, and~\citealp{kn:SecomandiYang2021}), which results from applying to the statistical measure the change of measure
\[
\frac{\prod_{j=i}^{\iota^\pi-1}\left(1-q^\pi_{j}(F_j)\Delta F_{j+1}\right)}{\E\left[\prod_{j=i}^{\iota^\pi-1}(1-q^\pi_{j}(F_j)\Delta F_{j+1})\mid \iota^\pi>i,F_i\right]}.
\]
Under the resulting measure, which can be signed, for each pair of stages $i$ and $j$ with $i<\iota^\pi$ and $j = i+1,\ldots,I-1$ the futures price process satisfies the martingale condition
\[
\mathbb{E}^{\mathrm{VO}(\pi)}\left[F_{\iota^\pi\wedge j}\mid \iota^\pi>i,F_i\right] = F_i,
\]
where $\wedge$ denotes minimum.
Define $D_{i,j}:=\prod_{i^\prime=i+1}^{j} D_{i^\prime}$ for each $i$ and $j\in\mathcal{I}$ with $j>i$.
For each stage $i\leq\iota^\pi$ and futures price $F_i$, we have
\begin{equation}
b^\pi_i(F_i) = C_i(F_i)1\{\iota^\pi=i\} + \mathbb{E}^{\mathrm{VO}(\pi)}\left[D_{i,\iota^\pi} C_{\iota^\pi}(F_{\iota^\pi})\mid \iota^\pi>i,F_i\right]1\{\iota^\pi>i\},
\label{eq:bchar}
\end{equation}
where $1\{\cdot\}$ is the indicator function.
In particular, the optimal initial bond position, $b^\pi_0(F_0)$, is $C_0(F_0)$ if $\iota^\pi=0$ and $\mathbb{E}^{\mathrm{VO}(\pi)}\left[D_{0,\iota^\pi} C_{\iota^\pi}(F_{\iota^\pi}) \mid \iota^\pi>0,F_0\right]$ if $\iota^\pi > 0$.

In complete markets the variance optimal martingale measure of any policy coincides with the unique risk neutral measure.
In this case expression~(\ref{eq:bchar}) evaluated on date $T_0$ for futures price $F_0$ reduces to risk neutral valuation (\citealp{kn:Shiryaev1999,kn:Duffie2001,kn:BinghamKiesel2004}, and~\citealp{kn:Shreve2004}): The term $b^\pi_0(F_0)$ is the date $T_0$ market value of the exercise policy $\pi$ for the option when the futures price is $F_0$.
If markets are incomplete then, using the terminology introduced by~\citet[p. 373]{kn:BertsimasEtAl2001}, in general this quantity is simply the minimal production cost of the option when using this policy.

\section{Exercise Policy Optimization}
\label{sc:opt}

If the hedged P\&L is determined according to the quadratic hedging optimal solution then it is tempting to look for an exercise policy that maximizes the option minimal production cost, taken to be an approximation for the option value (holders of financial options may not be interested in hedging, but managers of commodity and energy conversion assets modeled as real options may engage in this activity, as discussed, e.g., by~\citealp{kn:Pirrong2015}).
Doing so corresponds to solving
\[
\max_{\pi\in\Pi}b^\pi_{0}(F_0),
\]
which by~(\ref{eq:bchar}) can be equivalently written as
\begin{equation}
\max_{\pi\in\Pi}
\left\{C_0(F_0)1\{\iota^\pi=0\}+\mathbb{E}^{\mathrm{VO}(\pi)}\left[D_{0,\iota^\pi} C_{\iota^\pi}(F_{\iota^\pi})\mid \iota^\pi>0,F_0\right]1\{\iota^\pi>0\}\right\}.
\label{eq:general}
\end{equation}

An optimal solution to model (\ref{eq:general}) is meaningful if its associated variance optimal martingale measure is a risk neutral one, because in this case the value of this exercise policy is arbitrage free.
Otherwise, Example~\ref{ex:timeincons}, which is based on Example 4 in~\citet{kn:Secomandi2021}, shows that in general an optimal solution of model~(\ref{eq:general}) is an unappealing exercise policy that is costly to hedge, because it has maximal minimal production cost, rather than a policy that is optimal also under a risk neutral measure.



\begin{example}\label{ex:timeincons}
There are two dates ($I=2$).
The risk free interest rate is zero.
The futures price is \$3.20 on date $T_0$ and \$2.56, \$6.4, and \$16 with respective probabilities 0.05, 0.05, and 0.90 on date $T_1$.
Consider an American call option with strike price equal to \$3.
The optimal policy of model~(\ref{eq:general}) exercises this option on date $T_1$ when and only when the futures price is \$6.4.
This policy is intuitively unappealing.
Its minimal production cost is \$1.5286 (the weights of its corresponding variance optimal martingale measure for the date $T_1$ prices \$2.56, \$6.4, and \$16 are 0.6312, 0.4496, and $-$0.0808, respectively).
The optimal policy under any risk neutral measure exercises the option on the second date whenever the futures price exceeds the strike price.
It has a minimal production cost of \$0.4777.
This policy is naturally both more attractive and cheaper to hedge than the former one.
All the risk neutral measures are the ones that assign probability $\Pr^\mathrm{RN}\in(0,1/21)$ to the futures price \$16 on date $T_1$ and corresponding probabilities $5/6+5\Pr^\mathrm{RN}/2$ and $1/6-7\Pr^\mathrm{RN}/2$ to the futures prices \$2.56 and \$6.4 on this date.
Thus, the set of no arbitrage values for the optimal policy of model~(\ref{eq:general}) is $(\mbox{\$}0,\mbox{\$}1.7/3)\approx(0,\mbox{\$}0.5667)$ and the minimal production cost of this policy, \$1.5286, does not belong to it.
Moreover, the analogous set for the exercise policy that is optimal under any risk neutral measure is $(\mbox{\$}1.7/3,\mbox{\$}13/21)\approx(\mbox{\$}0.5667,\mbox{\$}0.6190)$, which excludes the minimal production cost of this policy, \$0.4777.
\end{example}



Example~\ref{ex:timeincons} indicates that in general one must refine model~(\ref{eq:general}) to obtain a reasonable exercise policy.
Optimal exercise policies determined under any risk neutral measure are time consistent: If this model were reformulated on a date following the initial one and the option had not been previously exercised then the corresponding residual part of the optimal policy obtained on the starting date would be optimal for this later optimization.
To formally state this property, let $\pi(i)$ be the part of policy $\pi$ corresponding to dates $T_i$ through $T_{I-1}$ and $\Pi(i)$ be the set of all such feasible policies ($\pi\equiv\pi(0)$ and $\Pi\equiv\Pi(0)$).
An optimal time consistent policy solves the following model for each date $T_i$ with $i\in\mathcal{I}$ and futures price $F_{i}\in\mathbb{R}_+$ for which the option is alive:
\begin{equation}
\max_{\pi(i)\in\Pi(i)}
\left\{C_i(F_i)1\left\{\iota^{\pi(i)}=i\right\}+\mathbb{E}^{\mathrm{VO}(\pi(i))}\left[D_{i,\iota^{\pi(i)}} C_{\iota^{\pi(i)}}(F_{\iota^{\pi(i)}})\mid \iota^{\pi(i)}>i,F_i\right]1\left\{\iota^{\pi(i)}>i\right\}\right\}.
\label{eq:generalstage}
\end{equation}
Let $\Pi^\mathrm{TC}$ be subset of the set of exercise policies $\Pi$ that satisfy time consistency.
The refinement of model~(\ref{eq:general}) that optimizes over this restricted set is
\begin{equation}
\max_{\pi\in\Pi^\mathrm{TC}}
\left\{C_0(F_0)1\{\iota^\pi=0\}+\mathbb{E}^{\mathrm{VO}(\pi)}\left[D_{0,\iota^\pi} C_{\iota^\pi}(F_{\iota^\pi})\mid \iota^\pi>0,F_0\right]1\{\iota^\pi>0\}\right\}.
\label{eq:refgeneral}
\end{equation}

The optimal policy of model~(\ref{eq:general}) obtained in Example~\ref{ex:timeincons} is time inconsistent.
Indeed, suppose that the realized futures price on date $T_1$ is \$16.
Formulating and solving model~(\ref{eq:generalstage}) for this date and price leads to a different decision compared to the one taken by the optimal policy obtained on date $T_0$: Exercising the option is now optimal.
In contrast, applying model~(\ref{eq:refgeneral}) to Example~\ref{ex:timeincons} yields the optimal exercise policy that one obtains under any risk neutral measure discussed in this example: Exercise on date $T_1$ when and only when the futures price equals \$6.4 or \$16.

Although model~(\ref{eq:refgeneral}) is an improvement compared to model~(\ref{eq:general}), it provides only a partial remedy to the outlined pitfalls.
First, in general the optimal value of the objective function of model~(\ref{eq:refgeneral}) is not an arbitrage free value.
In fact, the minimal production cost of the optimal policy of this model for Example~\ref{ex:timeincons} is \$0.4777, whereas its set of no arbitrage values is $(\mbox{\$}0.5667,\mbox{\$}0.6190)$---see Example~\ref{ex:timeincons}.
Second, model~(\ref{eq:refgeneral}) may not be able to yield a good exercise policy, as Example~\ref{ex:zerovalue} illustrates.

\begin{example}\label{ex:zerovalue}
Immediate exercise, which yields a cash flow equal to zero, is an optimal policy of both models~(\ref{eq:general}) and~(\ref{eq:refgeneral}) applied to the setting of Example~\ref{ex:timeincons} with the strike price set equal to \$7.
In contrast, under any risk neutral measure it is optimal to exercise the considered option on date $T_1$ when and only when the futures price is \$16.
The interval of no arbitrage values of this policy is $(0,\mbox{\$}3/7)$.
\end{example}

\section{Alternative Approach}
\label{sc:alternative}

\citet{kn:SecomandiYang2021} broaden to assets that generate streams of cash flows, which  include American options as a particular case, the proposal of~\citet{kn:Secomandi2021} that anchors quadratic hedging to the arbitrage free values of assets that can yield a single cash flow on a given date, like European options.
To apply this method in the setting of this paper, define the value of exercise policy $\pi$ under a given risk neutral measure as
\[
V^{\mathrm{RN},\pi}_0(F_0) :=\mathbb{E}^\mathrm{RN} \left[ 
D_{0,\iota^\pi}
C_{\iota^\pi}(F_{\iota^\pi})\mid F_0\right],
\]
where $\mathbb{E}^\mathrm{RN}$ is expectation under this measure (RN abbreviates risk neutral).
An optimal policy for the considered risk neutral measure solves
\begin{equation}
\max_{\pi\in\Pi} V^{\mathrm{RN},\pi}_0\left(F_0\right).
\label{eq:optpi}
\end{equation}
\citet[Chapter 8]{kn:Glasserman2004} and~\citet{kn:Detemple2005} discuss various approaches to solve this model.
For any exercise policy $\pi$, including one that is optimal for model~(\ref{eq:optpi}), quadratic hedging is performed in the conditional fashion
\[
\min_{\psi\in\Psi}\E\left[D^{-2}_{\iota^\pi,I-1}\cdot \left(C_{\iota^\pi}(F_{\iota^\pi})-V^{\psi}_{\iota^\pi}\right)^2\mid V^\psi_0 = V^{\mathrm{RN},\pi}_0(F_0), F_0\right],
\]
that is, in this model the initial risk free bond position, $V^\psi_0$, is forced to be equal to the risk neutral value of the considered exercise policy $\pi$, $V^{\mathrm{RN},\pi}_0(F_0)$.
This approach avoids the pitfalls of the procedure outlined in Section~\ref{sc:opt}.
It can be solved as discussed in Section~\ref{sc:opt} with $b^\pi_0(F_0)$ replaced by $V^{\mathrm{RN},\pi}_0(F_0)$.
 
\section{Conclusions}
\label{sc:concl}

Quadratic hedging is a practical approach to mitigate the risk embedded in option payoffs.
It gives rise to the variance optimal martingale measure.
One may be tempted to optimize an option  exercise policy under this measure.
Considering American options for simplicity of exposition, we demonstrate that pursuing this idea can lead to unappealing exercise policies.
This issue compounds the familiar drawback that using the variance optimal martingale measure to value an option can result in a value that is not arbitrage free.
An alternative and known approach that stays clear of these difficulties optimizes the option exercise policy under a risk neutral measure and uses the conditional version of quadratic hedging anchored to the value of this option.
Future research may quantify in the context of realistic applications the extent of the limitations of relying on the variance optimal martingale measure for option exercise policy optimization.

\section*{Acknowledgments}

This research is supported by NSF grant CMMI 1761742.
The author is a Faculty Affiliate of the Scott Institute for Energy Innovation at Carnegie Mellon University.
The author thanks Volodymyr Babich and Andrea Roncoroni for their comments on earlier drafts of this work.



\bibliographystyle{ormsv080}
\bibliography{mrchops,commodityStorage}

\end{document}